\documentstyle[aps,twocolumn,eqsecnum,epsfig,floats]{revtex}

\newcommand{\equ}[1]{(\protect\ref{#1})}
\newcommand{\reals}{{\rm I \mkern-2.5mu \nonscript\mkern-.5mu R}}

\newcommand{\ga}{\gamma}

\begin{document} 
\draft

\wideabs{
  
\title{Kinetic growth of field-oriented chains in dipolar colloidal
  solutions}

\author{M.-Carmen Miguel$^1$ and R. Pastor-Satorras$^2$}
\address{$^1$ Department of Physics\\
  Massachusetts Institute of Technology, Cambridge, Massachusetts
  02139\\
  $^2$ Department of Earth, Atmospheric, and Planetary Sciences\\
  Massachusetts Institute of Technology, Cambridge, Massachusetts
  02139\\
  \date{\today}}

\maketitle

\begin{abstract}
  Experimental studies on the irreversible growth of field-induced
  chains of dipolar particles suggest an asymptotic power-law behavior
  of several relevant quantities.  We introduce a Monte Carlo model of
  chain growth that explicitly incorporates the anisotropic diffusion
  characteristic of a rod-like object. Assuming a simple power-law
  form for the mean cluster size, $S(t) \sim t^z$, the results of our
  model are in good agreement with the experimental measurements of
  the dynamic exponent $z$.  Nevertheless, an alternative scenario,
  including logarithmic corrections to the standard power-law
  behavior, provides a better and more insightful interpretation of
  the anomalous dynamic exponent.  In contrast to some experimental
  findings, we do not observe any dependence of the exponents on the
  volume fraction of particles $\phi$. Finite-size effects are also
  explored by simulating very long time evolutions or highly
  concentrated systems.  Two different behaviors are found, namely,
  saturation and a crossover to a quasi one-dimensional regime.
\end{abstract}


}
\section{Introduction}
\label{sec:intro}

The kinetic properties of the irreversible aggregation of particles
has been the subject of great interest over the last decade
\cite{meakin93,vicsek92,jullien87}. In particular, several
experimental and theoretical studies, as well as computer simulations
have been devoted to understanding the behavior of colloidal suspensions
of dipolar particles, the so-called {\em electrorheological} and {\em
  magnetorheological fluids}
\cite{proceedings93,fraden89,bossis90,fermigier92,promislow95,%
  davies86,miyazima87,helgesen88}.  After applying an external
electric or magnetic field, these suspensions experience a dramatic
change in their rheological properties---i.e., a notable increase in
the viscosity---which renders them particularly interesting systems
from a technological point of view.

The change in the rheological properties of dipolar suspensions upon
the action of an external field, is mainly due to the aggregation of
the colloidal particles, which form clusters of macroscopic size.
These are usually linear chains---rods---oriented along the direction
of the applied field, although for high enough concentrations of
dipolar particles more complex structures may arise
\cite{wang94,jund95}. The overall spatial arrangement of the
aggregates is very effective in hindering the fluid flow, conferring
the suspension a solid-like texture.

Multiple applications have been envisaged for these materials:
lubricants, dampers, heat and light transmission devices, etc.
Nevertheless, their eventual manufacture still has to overcome
different drawbacks, as for instance, the sedimentation of the
clusters.  At high fields, the aggregation is irreversible, the chains
grow longer and never break as long as the field is present. Thus, the
study of the kinetics of formation of the aggregates turns out to be
an issue of practical importance.

Due to its irreversible character, studies have most often focused on
the dynamic properties of the distribution of clusters, following the
approach developed to deal with irreversible cluster-cluster
aggregation models (see Refs.~\cite{vicsek92,jullien87}, and
references therein). Within this framework, the quantitative features
of the aggregation are commonly described in terms of the {\em
  cluster-size distribution} as a function of time, $n_s(t)$, which is
defined as the number of clusters of size $s$, per unit volume,
present in the system at time $t$. Other relevant quantities are the
density of clusters at time $t$, $n(t)=\sum_s n_s(t)$, and the
{\em mean cluster size} $S(t)$, defined by
\begin{equation}
S(t)=\frac{\sum_s s^2 n_s(t)}{\sum_s s n_s(t)}.
\end{equation}
Experimental results and computer simulations of different
cluster-cluster aggregation models \cite{vicsek92,jullien87} show that
the asymptotic behavior at large times of the mean cluster size and
the total number of clusters is a power law:
\begin{equation}
S(t) \sim t^{z}, \qquad \qquad n(t) \sim t^{-z}.
\label{powerlaws}
\end{equation}
The exponent $z$ is the so-called {\em dynamic exponent}. In
general, $z$ depends on the dimension $d$ of the space, as well as on
the nature of the aggregation process.  Another relevant feature of
the aggregation dynamics is that the characteristic quantities defined
above are related through the {\em dynamic scaling hypothesis}
\cite{vicsek84}:
\begin{equation}
n_s(t) = s^{-2} F[s/S(t)],
\label{dynamicscaling}
\end{equation}
where $F(x)$ is a certain scaling function, independent of $t$ and
$s$. The validity of this relation has been systematically verified in
experiments and computer simulations.

The dynamics of cluster-cluster aggregation can be theoretically
described in terms of the Smoluchowski equation \cite{smoluchowski16}.
Smoluchowski proposed a kinetic equation describing the temporal
evolution of the cluster-size distribution $n_s(t)$:
\begin{equation}
  \frac{d n_s(t)}{d t} = \sum_{i+j=s} K(i, j) n_i(t) n_j(t) -  n_s(t)
  \sum_{i=1}^\infty K(s, i)  n_i(t).
  \label{smoluchow}
\end{equation}
Here $K(s,s')$ is the reaction kernel, giving the rate at which
clusters of size $s$ join clusters of size $s'$ to form clusters of
size $s+s'$.  This equation is an example of a mean-field-type theory
in which fluctuations are neglected. Moreover, its range of
applicability is limited to low concentrations, where the assumption
of exclusively binary collisions is valid \cite{smoluchowski16}. The
results of this theory are expected to hold for dimensions higher than
the {\em upper critical dimension} ($d_c=2$ in this case
\cite{kang86}), above which fluctuations become irrelevant.
Assuming that the reaction kernel is a homogeneous function of degree
$\ga$, that is,
\begin{equation}
  K(b s, b s') \equiv b^\ga K(s,s'), 
  \label{kernelscal}
\end{equation}
it is possible to prove  the relations \equ{powerlaws}
and \equ{dynamicscaling} \cite{jullien87}, with a dynamic exponent given by
\begin{equation}
z=\frac{1}{1-\ga}.
\label{smoluexponent}
\end{equation}

Most of the experimental and numerical studies of the anisotropic
aggregation of {\em rod-like clusters} presented so far
\cite{fraden89,fermigier92,promislow95,miyazima87,maas96} seem to
indicate the same type of power-law for the asymptotic behavior of the
mean cluster size $S(t)$.  For instance, Fraden {\em et al.}
\cite{fraden89} reported a value of $z=0.60$.  Promislow {\em et al.}
\cite{promislow95}, on the other hand, measured a value of $z$ ranging
between $0.50$ and $0.75$ for different values of the concentration
and the dipolar interaction strength $\lambda$ (to be defined later
on).  The interpretation of these results relies on Smoluchowski's
assumptions.

In Smoluchowski's solution \cite{smoluchowski16}, the reaction kernel
is the product of an effective diameter---the collision cross section
of two clusters---$(R_s + R_{s'})$, times an effective relative
diffusion coefficient $(D(s) + D(s'))$, where $R_s$ and $D(s)$ are the
radius of influence and the diffusion coefficient of a cluster of mass
$s$, respectively \cite{footn3}. By analogy with the Stokes-Einstein
relation for a single spherical particle of diameter $a$ moving in a
liquid, $D\propto a^{-1}$ \cite{batchelor67}, the effective diffusion
coefficient is given by a power law of the cluster size,
\begin{equation}
D(s) \sim s^{\ga}.
\label{diffusion}
\end{equation}

Miyazima {\em et al.}  \cite{miyazima87} proposed this type of
power-law kernel for rod-like clusters of dipolar particles. One can
estimate the effective cross section---the radius of influence---of a
single magnetic particle, $R_1$, by comparing dipolar ($U_d(r)\propto
m^2/r^3$) and thermal ($k_BT$) energies.  If we introduce the
dimensionless parameter $\lambda= m^2/(a^3 k_B T)$, where $m$ is the
magnetic moment and $a$ the diameter of a dipolar particle, the
effective cross section can be expressed as $R_1/a\sim \lambda^{1/3}$.
The detailed form of the dipolar interaction is not expected to modify
the asymptotic behavior of the aggregation dynamics. Thus, roughly
speaking, outside a spherical region of radius $R_1$ the relative
motion is mainly diffusive, and only when one particle enters the
sphere of influence of another they stick irreversibly.

At low concentrations, it is reasonable to assume that rod-like
clusters of dipolar particles essentially aggregate tip-to-tip.
Miyazima {\em et al.} argued that, under such conditions, the cross
section of a chain cannot depend on its total length \cite{footn}, and
it is possible to approximate $R_s\sim R_1$. The only source of
dependence on the cluster size should thus come from the effective
diffusion coefficient, and the reaction kernel is approximately given
by $K(s,s') \sim s^{\ga} + s'^{\ga}$. This form fulfills the
homogeneity condition \equ{kernelscal} and yields therefore a dynamic
exponent given by Eq.~\equ{smoluexponent}.

Following this approach to the problem, the value of $\ga$ is the key
parameter in order to interpret the experimental data
\cite{fraden89,fermigier92,promislow95}.  The dynamic exponent $z$ is
numerically computed from a log-log plot of the mean chain length
$S(t)$ as a function of time. The value obtained is then associated
through Eq.~\equ{smoluexponent} to a particular value of $\ga$.  In
the absence of hydrodynamic interactions among the spheres in a chain,
the effective mobility is inversely proportional to its length.
Consequently, one would expect $\ga$ to be equal to $-1$ which,
according to~\equ{smoluexponent}, corresponds to $z=1/2$. In general,
this value of $z$ deviates from the ones measured experimentally.  A
simple way of interpreting this difference is to consider an {\em ad
  hoc} value of $\ga$ different from $-1$, which phenomenologically
accounts for a more realistic mobility of a rod-like particle
\cite{happel86}.

In this paper we propose an alternative mechanism to explain those
discrepancies in the measured value of $z$. The core of our proposal
is a different view of the mobility of a cluster of mass $s$.  Fraden
{\em et al.}  \cite{fraden89} already pointed out that hydrodynamic
interactions should {\em increase} the effective mobility of a cluster
of mass $s$, making it larger than the mobility of a collection of $s$
independent particles. The latter approximation, stating $D(s) \sim
s^{-1}$, is strictly valid for clusters of {\em hydrodynamically
  noninteracting} spheres \cite{rouse53}.  However, this is too strong
an assumption in the case under consideration, where the magnetic
spheres form relatively rigid chains.  In this case, it is well-known
that the hydrodynamic drag for the translational motion is {\em
  anisotropic} \cite{happel86,doi86}.  The drag coefficients along the
direction parallel to the axis of the chain and in the perpendicular
directions, $\xi_\parallel$ and $\xi_\perp$ respectively, are
approximately given by \cite{doi86}
\begin{equation}
  \xi_\parallel = 2 \pi \eta a \frac{s}{\ln (s)} \qquad  \mbox{\rm
    and} \qquad  \xi_\perp = 2 \xi_\parallel,
  \label{eq:mobilidad}
\end{equation}
where $\eta$ is the viscosity of the solvent and $s$ is the number of
particles composing the cluster---its mass. The
mobility of a cluster is defined as the inverse of the drag
coefficient; hence, the constants  $D_\parallel$ and $D_\perp$,
characterizing the diffusion parallel and perpendicular to the rods'
axis, are given by \cite{footn3}
\begin{equation}
  D_\parallel=\frac{k_B T}{\xi_\parallel} \sim \frac{\ln(s)}{s}
  \qquad \mbox{\rm  and} \qquad  D_\perp=\frac{k_B T}{\xi_\perp} =
  \frac{D_\parallel}{2}.
  \label{slenderbody}
\end{equation}
That is to say, hydrodynamic interactions generate anisotropic
diffusion coefficients, exhibiting logarithmic corrections to the
previously considered power laws. In order to investigate the effect
of the new diffusivities \equ{slenderbody} on the dynamics of the
process, we have introduced them in a Monte-Carlo \cite{binder86}
computer model for the cluster-cluster aggregation of rod-like
particles.

By simply considering anisotropic diffusivities with logarithmic
corrections, and assuming that $S(t) \sim t^z$, we are able to recover
a dynamic exponent $z$ in good agreement with experimental results.
However, a simple heuristic argument suggests a different functional
form, namely, a power-law with logarithmic corrections.

Let us assume that all clusters in the suspension have the same
average length $S$. The average separation $\overline{R}$ between
neighboring clusters in a suspension with initial volume fraction
$\phi$---the initial density of dipoles---can then be estimated to be
\begin{equation}
  \label{eq:mindis}
  \overline{R}\simeq \left(\frac{S}{\phi}\right)^{1/d}.
\end{equation}
In $d=1$, only two neighboring clusters are able to aggregate. In
general, this is the most likely event in low dimensions. In high
dimensions, however, and due to the diffusive nature of their
movement, any two clusters are equally likely to join, irrespective of
their relative distance, as they are quite invisible to each other. A
mean-field type of behavior is thus expected, in the sense that only
the density of clusters present $n(t)$ is relevant, and not their
spatial arrangement.

As the movement of the clusters is essentially diffusive, the average
displacement of a cluster of mean size $S$ after the time interval $t$
is given by

\begin{equation}
  \label{eq:argumento}
  \langle R^2 \rangle\propto D(S) \ t,
\end{equation}
where, as indicated in Eq.\equ{slenderbody}, $D(S)\sim\ln(S)/S$.
Aggregation of two clusters will occur after a characteristic time $T$
has been elapsed; a time interval long enough for the clusters to
cover their relative separation.

For a low-dimensional system, the characteristic time $T$ is that
required to cover the distance $\overline{R}$ separating near
neighbors, i.e.,
\begin{equation}
  \label{eq:argld}
   T\propto\frac{\overline{R}^2}{D(S)}\propto\frac{1}{\phi^{2/d}}
   \frac{S^{(2+d)/d}}{\ln(S)}.
\end{equation}
On the other hand, in higher dimensions there is no such
characteristic length scale and, consequently, there is no reason to
expect that the previous expression holds. However, we can argue that
a cluster browsing a volume of order unity will encounter $n(t)$
clusters available to join to. The time needed for a cluster to cover
such space will be thus proportional to $n T$, where, as above, $T$ is
the characteristic time of a {\em single} aggregation event. We have
then
\begin{equation}
  \label{eq:arghd}
  T\propto\frac{1}{n}\frac{1}{D(S)}\propto\frac{1}{\phi}\frac{S^2}{\ln(S)}, 
\end{equation}
where we have used that $n\sim \phi / S$. Note that both
estimations yield the same result at  $d=2$, suggesting that the upper
critical dimension of the problem is $d_c=2$ \cite{kang86}.

The inspection of Eqs. \equ{eq:argld} and \equ{eq:arghd} suggests that
the functional dependence of the mean cluster size with time takes the
form, for $d\leq d_c$,
\begin{equation}
  \label{eq:mfld}
  \frac{S}{[\ln(S)]^\zeta} \sim (t\phi^{2/d})^\zeta,
\end{equation}
where $\zeta=d/(2+d)$; and for $d\geq d_c$,
\begin{equation}
  \label{eq:mfhd}
  \frac{S}{[\ln(S)]^{1/2}} \sim (t\phi)^{1/2}.
\end{equation}
Note that within this approach, we obtain logarithmic corrections to
the behavior reported in Ref.~\cite{miyazima87} for $\gamma=-1$ (see
Eq.\equ{smoluexponent}). These corrections to the usual asymptotic
behavior in {\em isotropic} cluster-cluster aggregation could explain
the anomalous dynamic exponent found for the aggregation of
anisotropic rod-like clusters. The results of our numerical
simulations exhibit a surprisingly good fit to the theoretical
predictions \equ{eq:mfld} and \equ{eq:mfhd}, better than to the naive
power law behavior.

We have structured our paper as follows. In Section~\ref{sec:model} we
describe the technical details of our algorithm.
Section~\ref{sec:d=2} deals with the properties of rod-like
aggregation in two dimensions.  In particular, we study the mean
cluster size $S(t)$ for a variety of values of the initial volume
fraction $\phi$. For small to moderate values of $\phi$ and not very
long execution times, we recover, assuming a pure power law form for
$S(t)$, values of $z$ in agreement with some of the experimental
findings. Importantly, in opposition to some claims in the literature
\cite{promislow95}, we do not observe any dependence of $z$ on the
volume fraction.  On the other hand, we observe that the same sets of
data can be fitted with higher accuracy to our predicted power law
with logarithmic corrections. By allowing very high volume fractions
or large execution times, we find that our model crosses over to two
different behaviors, namely, saturation and a quasi one-dimensional
regime. We believe that the discrepancies with the experimental
results in Ref.~\cite{promislow95} might be due to a combination of
this type of finite-size effects and logarithmic corrections. In
Sec.~\ref{sec:d=3} we extend our model to three dimensions. We do not
observe significant variations with respect to the two-dimensional
aggregation. Finally, our conclusions are presented in
Sec.~\ref{sec:conclu}.

\section{Computer model}
\label{sec:model}

In our model we consider the irreversible aggregation of rigid
rod-like clusters in $\reals^d$, with $d=2$ and $3$.  Clusters are
oriented along the $Z$ axis.  Simulations start at $t=0$ with a random
distribution of $N_0$ monomers (spherical particles of diameter $a=1$)
in a box of volume $V$ with periodic boundary conditions. The initial
volume fraction of monomers is defined by $\phi=N_0/V$.  We have
simulated a wide range of volume fractions $\phi$ ($0.001$ to $0.1$),
and some particular high values (up to $0.5$).  Clusters diffuse
performing a free off-lattice random walk; that is, despite the
dipolar interactions, we assume that the temperature in the system is
high enough to provide a mainly diffusive character to the aggregation
dynamics.  Nevertheless, dipolar interactions are predominant at short
distances. Therefore, when two clusters come close enough to each
other, they stick irreversibly.

In the Monte-Carlo algorithm, clusters are selected and moved a
distance equal to one diameter $a$ in a direction chosen randomly from
a certain probability distribution.  The movement is performed
rigidly, preserving the orientation of the rod along the $Z$ axis.
When two clusters come within a distance ``tip-to-tip'' of one
diameter (that is, when the distance between any two of their
respective ends is less than or equal to $a$) they join, forming one
single rod of mass---number of particles---equal to the sum of the
masses of the colliding clusters. When two clusters approach
``side-to-side'' they repel each other.  This is implemented by
rejecting all possible movements leading to a ``side-to-side'' overlap
of clusters.

In fact, this procedure is equivalent to considering a radius of
influence $R_s=a$, and, according to the discussion in the previous
section, it mimics a dipolar interaction of strength $\lambda\sim 1$.
Different values of $\lambda$ could be in principle simulated by
joining clusters irreversibly when they come within a distance
$\sim\lambda^{1/3}$.  Values of $\lambda$ between $1$ and $30$ (that
is, radius of influence between $1$ and $3$), within the range
reported in experimental investigations, provide analogous asymptotic
results at large times, differing only in the transient regimes.

The effect of the anisotropic diffusivities \equ{slenderbody} on the
algorithm is taken into account in two steps: First, in the selection
of the next cluster to be moved, and second, in the prescription for
the direction of its tentative movement.  In the first step, we choose
a cluster among all present at a given time step, with a
probability $\rho(s)$ proportional to their diffusivity. According to
Eq.~\equ{slenderbody}, we have
\begin{equation}
  \rho(s) \sim \frac{\ln(s)}{s}.
\end{equation}
However, given that the expression \equ{slenderbody} is only suitable
to describe the motion of a {\em slender} body ($s\gg1$), the previous
expression would be inappropriate for small $s$. Therefore, we
actually select the cluster with a probability proportional to the
corrected diffusivity
\begin{equation}
  \rho(s) \sim \tilde{D}_C(s)=\left\{ \begin{array}{ll} 
      \ln(s_0) /s & \qquad \mbox{\rm for } s  < s_0\\ 
      \ln(s) / s & \qquad \mbox{\rm for } s \geq s_0 
    \end{array} \right. .
  \label{effectivediff}
\end{equation}
Here we are assuming that small clusters $(s<s_0)$ diffuse as if
composed of hydrodynamically independent particles.  We have selected
a cut-off mass $s_0=3$; different values were also tested, yielding
comparable results.

We discuss now the method employed to sample the clusters. In
cluster-cluster aggregation simulations, the sampling with a
probability proportional to the mobility is usually performed
according to the ``rejection'' algorithm (\cite{vicsek92,meakin85} and
Ref.~\cite{bratley87}, p. 151): A cluster is selected among all the
present at a given time, with uniform probability. Then, a random
number $\eta$, uniform in the interval $[0,1]$, is drawn. The cluster
selected is accepted for movement if $\eta < D(s)/D_{\rm max}$, where
$D(s)$ is the diffusion coefficient of the cluster considered, and
$D_{\rm max}$ is the maximum diffusivity of all the clusters present.
Otherwise, the cluster is rejected, and both selection steps are
repeated. In our simulations, however, we have chosen to implement a
different sampling algorithm, the ``alias'' method for discrete
distributions (see Ref.~\cite{bratley87}, p. 158, for a detailed
description).  Briefly, the algorithm works as follows: A cluster $C$,
of size $s$, is selected uniformly among all clusters.  Then, with a
certain ``aliasing'' probability $p(s)$, the cluster is replaced by
its alias $C' = {\cal F} [C]$, of mass $s'$.  The probability $p(s)$
is chosen so that clusters with small $\tilde{D}_C(s)$ (low
probability of being selected) are frequently mapped to clusters with
higher $\tilde{D}_C(s')$. The ``alias'' method is more costly in
computer time; however, it provides a more accurate sampling of the
probability density $\rho(s)$.

With the procedure described above, we implement the mass dependence
of the mobility \equ{slenderbody}.  However, we also have to take into
account its {\em anisotropy}.  Given Eq.~\equ{slenderbody}, it is
twice more likely for any rod to move along its axis that along any
other perpendicular direction. We implement this fact in a second step
by selecting the direction of the trial movement from a probability
distribution fulfilling this same anisotropy. Let us define the
azimuthal angle $\theta$ with respect to the axis of the rod.  Then,
in $d=2$, the direction for the tentative movement of a chain of
length $s \geq s_0$ is selected at random from the probability density
\begin{equation}
  P(\theta)=\frac{\sqrt{2}}{2 \pi} \frac{1}{1+\sin^2 \theta},
  \label{probangle-2d}
\end{equation}
with $0<\theta<2\pi$. In $d=3$ we use instead the density
\begin{equation}
  P(\varphi, \theta)=\frac{1}{2\pi} \frac{\sqrt{2}}{\ln(3 + 2 \sqrt{2})}
  \frac{1}{1+\sin^2 \theta},
  \label{probangle-3d}
\end{equation}
with $0<\varphi<2\pi$ and $0<\theta<\pi$.  These distributions ensure
the necessary anisotropic condition $P[\theta=0, \pi]/ P[\theta=\pi/2,
3\pi/2] = 2$, while being continuous in both angles, and easily
simulated numerically. For clusters of mass $s<s_0$ we assume that the
diffusion is isotropic.  Therefore, in this case we choose a random
direction for the tentative movement from a uniform distribution.

The final implementation of our algorithm runs as follows: (i) Each
time step we select at random a cluster of mass $s_i$ according to the
probability density \equ{effectivediff} and a direction $\theta_i$,
according to \equ{probangle-2d} [a pair $(\theta_i, \varphi_i)$,
according to \equ{probangle-3d}, in three dimensions] for $s_i \geq
s_0$; in case that $s_i<s_0$, the direction is drawn from a uniform
distribution. (ii) The cluster is moved a distance $a$ in the selected
direction, and its position with respect to the neighboring chains is
analyzed. (iii) If it intersects ``side-to-side'' with another chain,
the movement is rejected; otherwise, it is accepted. If the cluster
intersects ``tip-to-tip'' with another chain, they join and form a
single cluster.  (iv) Finally, the time is incremented by an amount
\begin{equation}
  \Delta t=\frac{1}{N(t)\tilde{D}_C(s_i)},
  \label{eq:deltatime}
\end{equation}
where $N(t)$ is the total number of clusters present at the
corresponding time $t$, and $\tilde{D}_C(s_i)$ is the effective
diffusivity of the selected cluster, as given by
Eq.~\equ{effectivediff}. This choice of the time step effectively
reproduces the real dynamics of the system.

\section{Aggregation in \lowercase{$d=2$}}
\label{sec:d=2}

In our simulations in $d=2$ we have mainly considered systems with
volume ranging from $256 \times 128$ to $1024 \times 512$, and initial
number of particles between $524$ and $67109$. This corresponds to
volume fractions ranging between $\phi\sim0.001$ and $\phi\sim0.1$. We
have also considered the extreme case of high concentration
($\phi\sim0.5$) in order to investigate a possible crossover to a
one-dimensional regime. In most cases, averages were performed over
$100$ to $500$ simulations.

\subsection{Isotropic diffusion}

Firstly, in order to check our algorithm, we have simulated the
aggregation according to the prescription given in
Ref.~\cite{miyazima87}, i.e., the diffusivity of a rod is isotropic
and inversely proportional to its mass, $D(s)\sim s^{-1}$.

\begin{figure}[t]
  \centerline{\epsfig{file=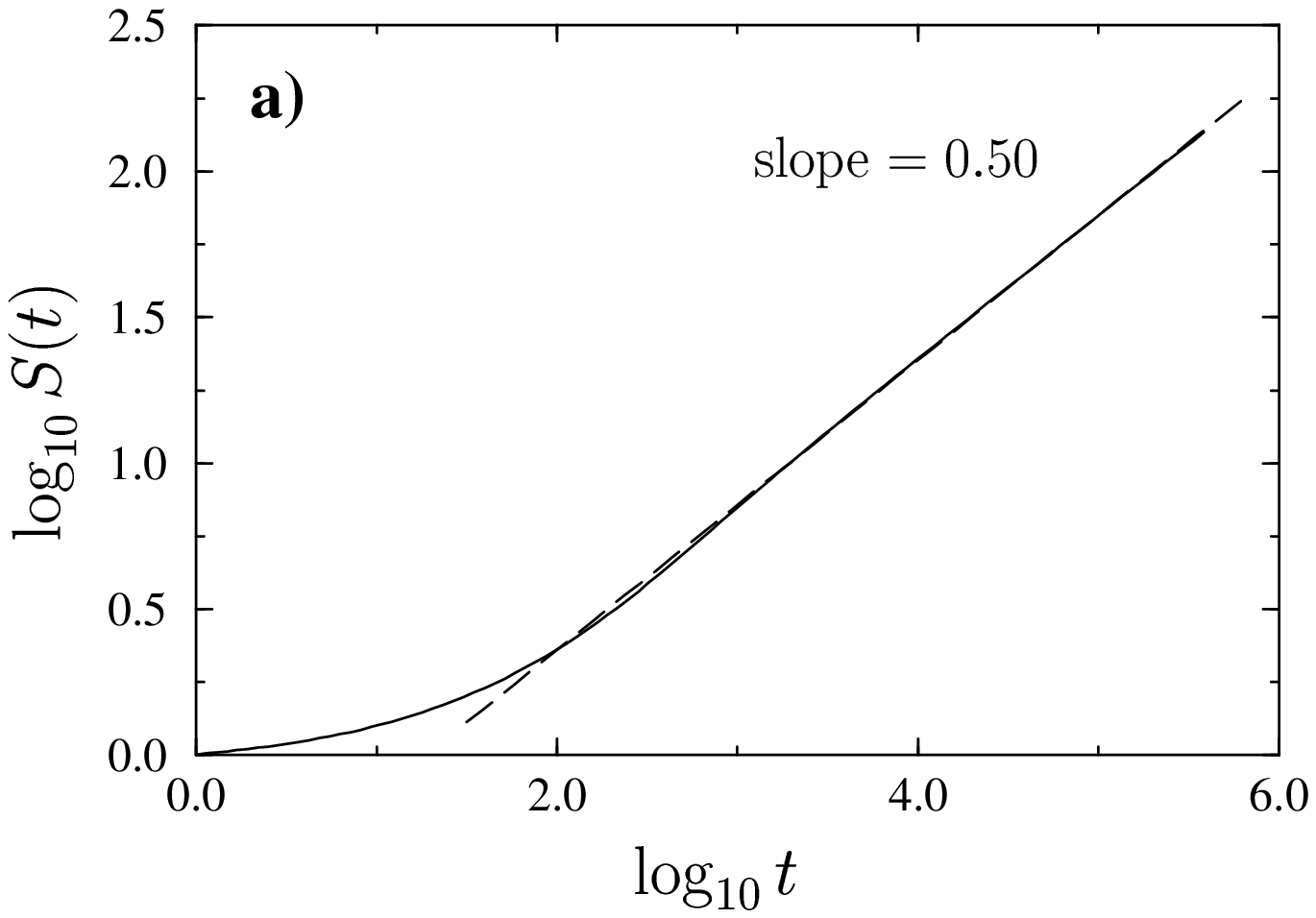, bbllx=131,
      bblly=375, bburx=540, bbury=660, width=8.5cm}}
  \centerline{\epsfig{file=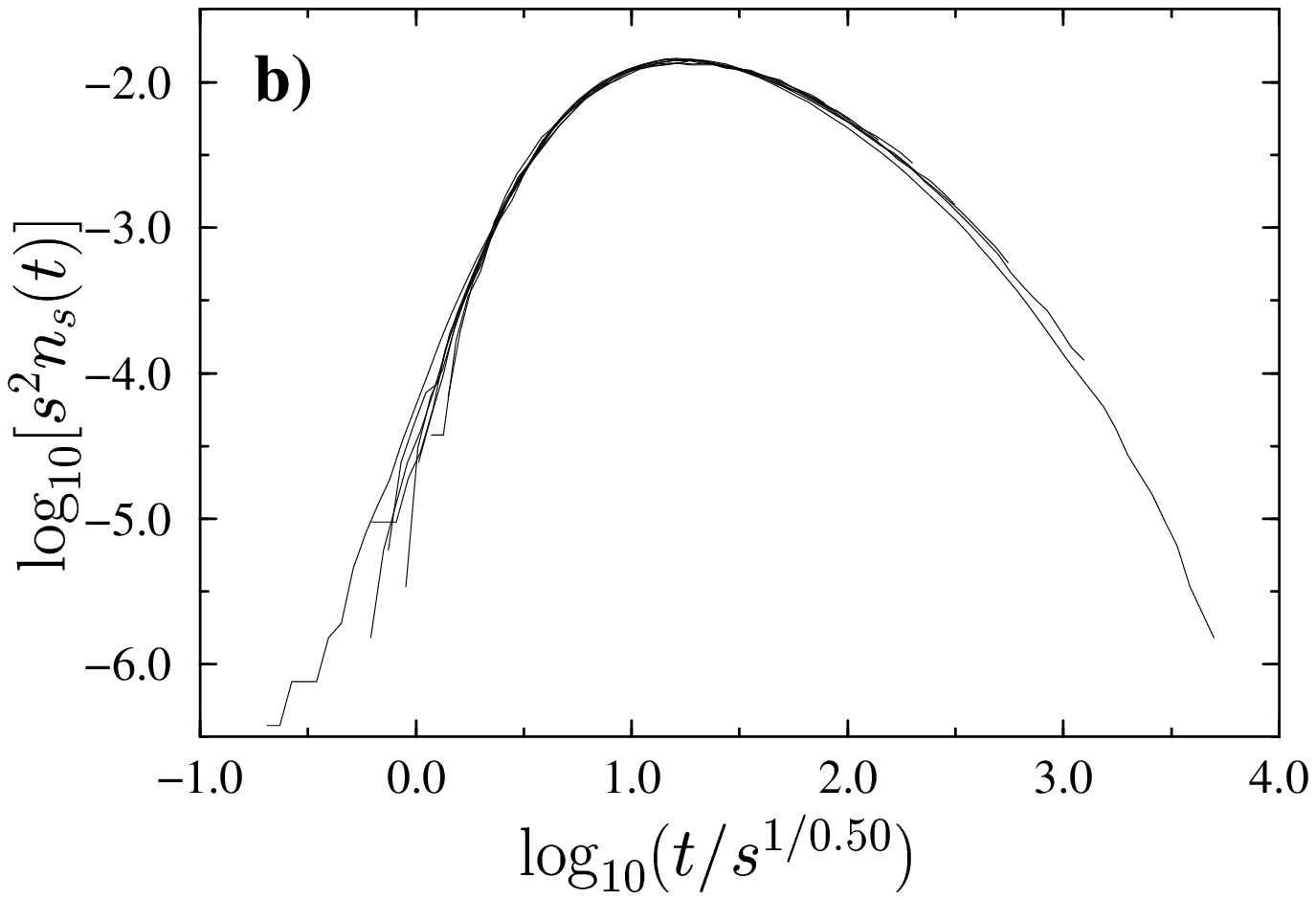, bbllx=131,
      bblly=375, bburx=540, bbury=660, width=8.5cm}}
  \caption{Aggregation with cluster diffusion that is isotropic and
    proportional to $s^{-1}$.  Initial concentration
    $\phi=0.019$. a) Mean cluster size as a function of time.  b)
    Finite-size scaling of the cluster size distribution as function
    of time, for fixed lengths $s=10, 20, 30, 40, 50, 60, 70,$ and
    $80$.}
  \label{fig:test}
\end{figure}

In Fig.~\ref{fig:test}(a), we represent the mean cluster size $S(t)$
obtained from simulations of a system of size $1024\times512$, with an
initial number of particles $N_0=10000$.  These values correspond to
an initial volume fraction $\phi=0.019$.  We observed a clear power
law regime covering more than three decades in time. The least-squares
fitting of the curve yields an exponent $z=0.50$. As expected, this
result matches the dynamic exponent predicted by \equ{smoluexponent}
for $\gamma =-1$. In Fig.~\ref{fig:test}(b) we have tested the
finite-size scaling relationship \equ{dynamicscaling} by plotting
$s^2n_s(t)$ as a function of the rescaled time $t/s^{1/z}$ for
different values of the cluster size $s$.  The best collapse of the
plots is obtained for a value of $z=0.50$, which is again in agreement
with \equ{smoluexponent}.

\subsection{Anisotropic diffusion: dilute regime}

\begin{figure}[t]
  \centerline{\epsfig{file=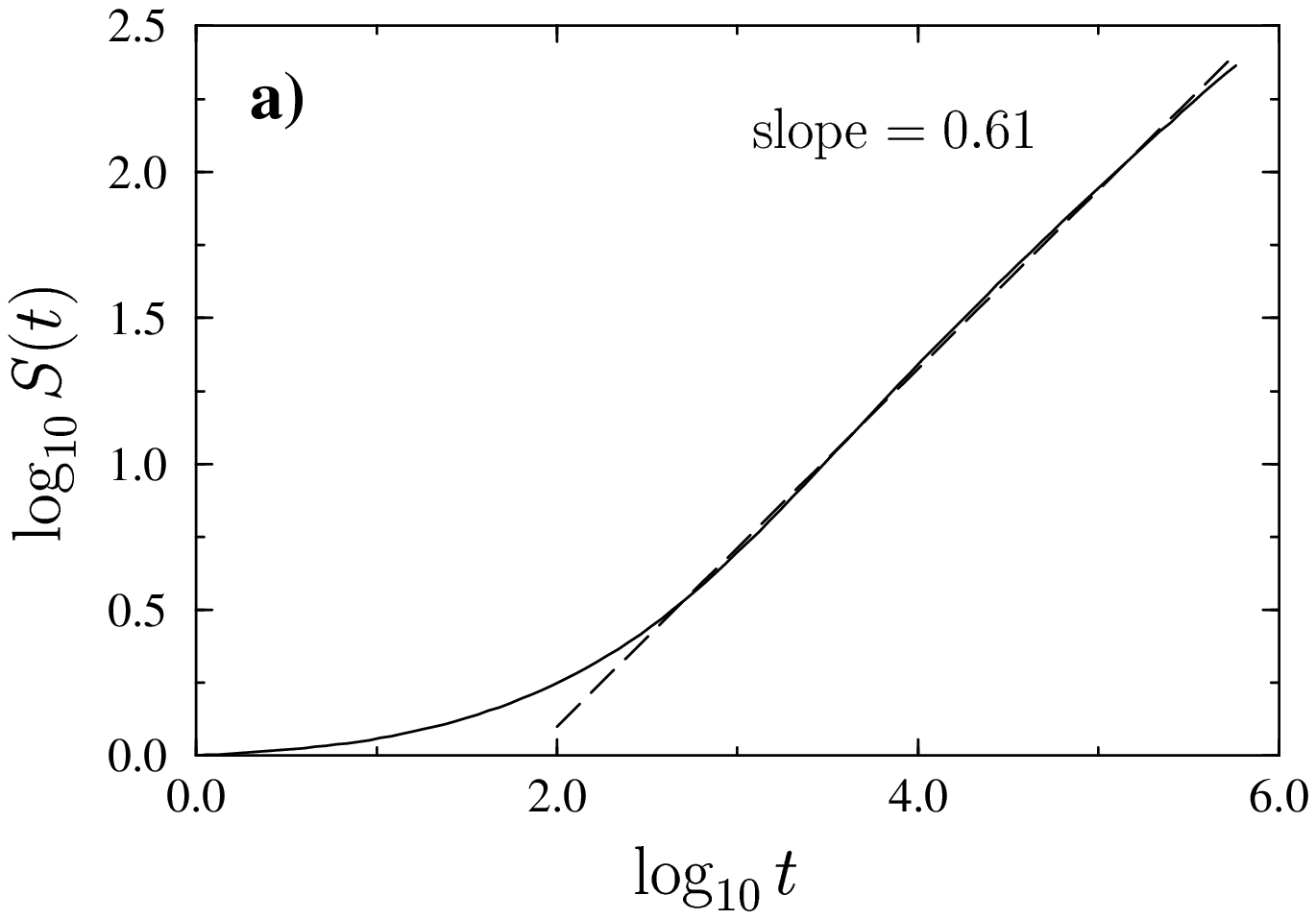,
      bbllx=131, bblly=375, bburx=540, bbury=660, width=8.5cm}}
  \centerline{\epsfig{file=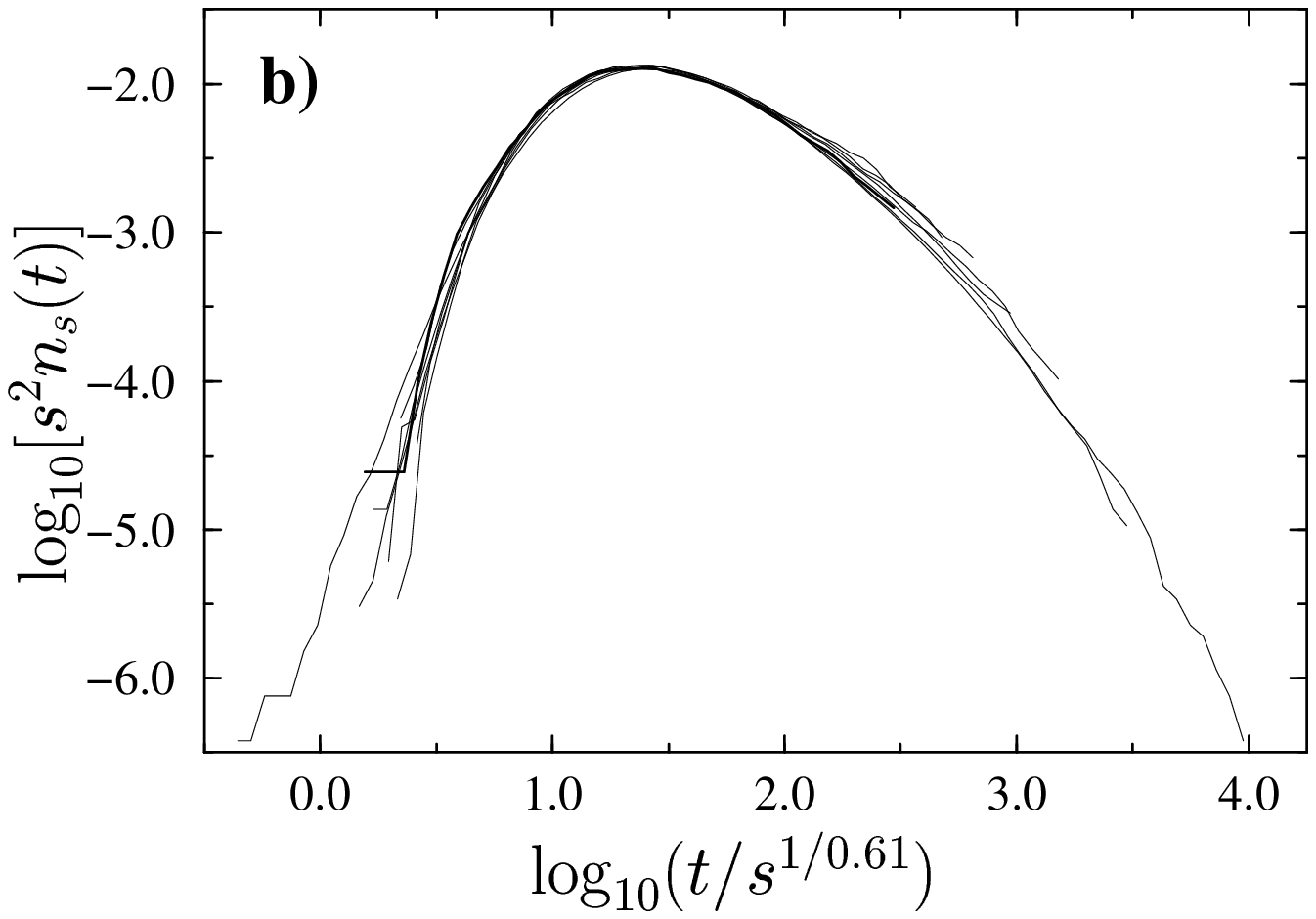,
      bbllx=131, bblly=375, bburx=540, bbury=660, width=8.5cm}}
  \caption{Aggregation with cluster diffusion that is anisotropic and
      proportional to $\ln(s)/s$, as defined in Eqs.~\equ{effectivediff}
      and~\equ{probangle-2d}. Initial  concentration
      $\phi=0.009$, as in Ref.~\protect\cite{fraden89}.
    (a) Mean cluster size as a function of
    time.  (b) Finite-size scaling of the cluster size distribution as
    function of time, for fixed lengths 
    $s=10, 20, 30, 40, 50, 60, 70,$ and $80$.}
  \label{fig:2d-hidroaniso}
\end{figure}

We now discuss the results obtained by implementing into our algorithm
the anisotropic diffusion prescribed in $d=2$ by
Eqs.~\equ{effectivediff} and~\equ{probangle-2d}. First of all,
motivated by the results in Ref.~\cite{fraden89}, we have duplicated
the parameters in that experiment, i.e., we have chosen a system of
volume $1024\times512$ and an initial number of particles $N_0=5000$.
These conditions yield a volume fraction $\phi\sim0.01$, as in the
experiments.  Figure~\ref{fig:2d-hidroaniso}(a) shows the mean cluster
size as a function of time, in a double-logarithmic plot. The behavior
of $S(t)$ at late times can be fitted to a power law extending close
to three orders of magnitude. A least-squares fitting provides an
exponent $z=0.61$, in complete agreement with the experimental
findings in \cite{fraden89}.  Similarly, in
Fig.~\ref{fig:2d-hidroaniso}(b) we represent the finite-size scaling
hypothesis \equ{dynamicscaling}. The good collapse shown in this plot
corroborates the value of the dynamic exponent $z=0.61$.

\begin{figure}[t]
  \centerline{\epsfig{file=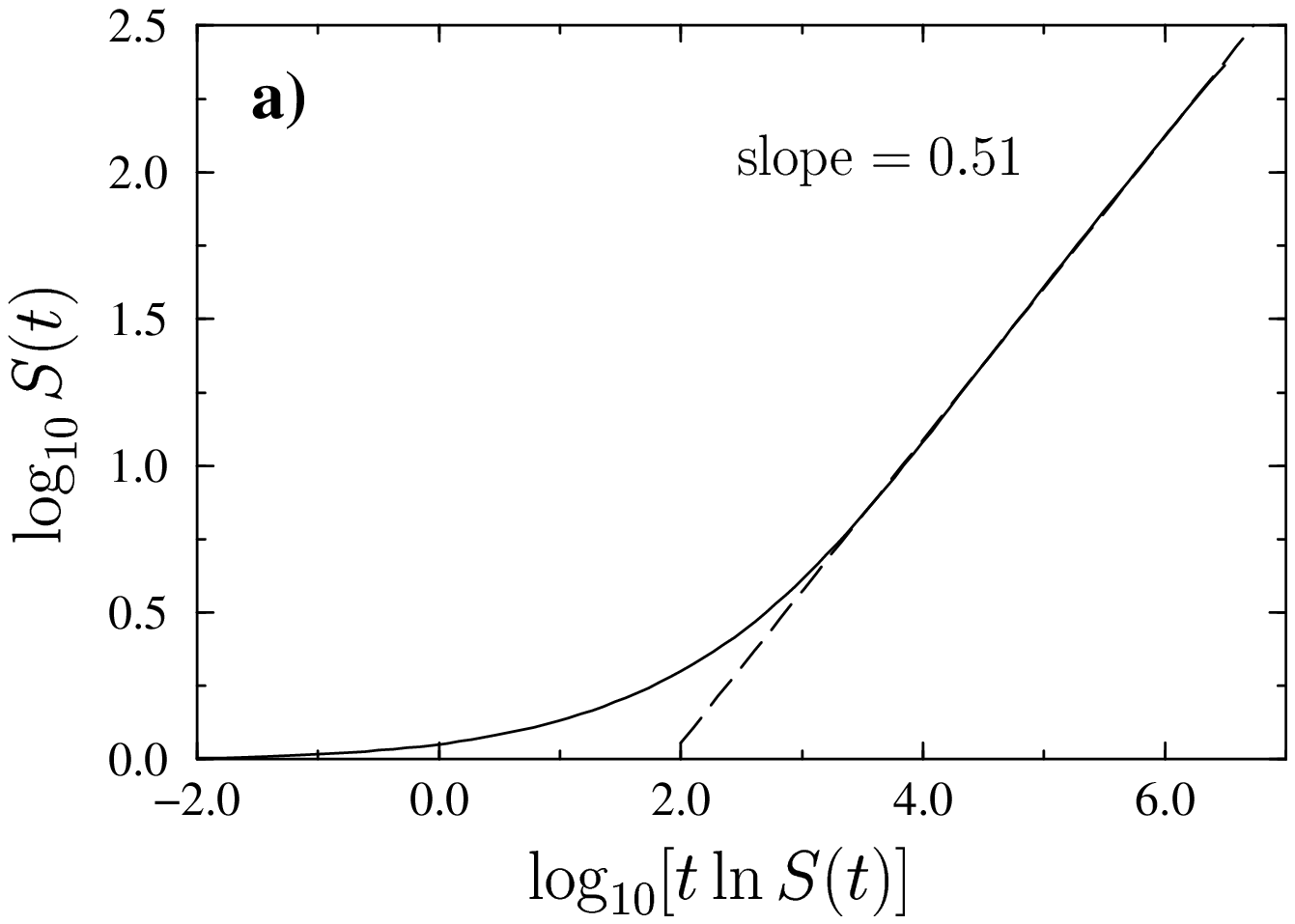,
      bbllx=131, bblly=375, bburx=540, bbury=660, width=8.5cm}}
  \centerline{\epsfig{file=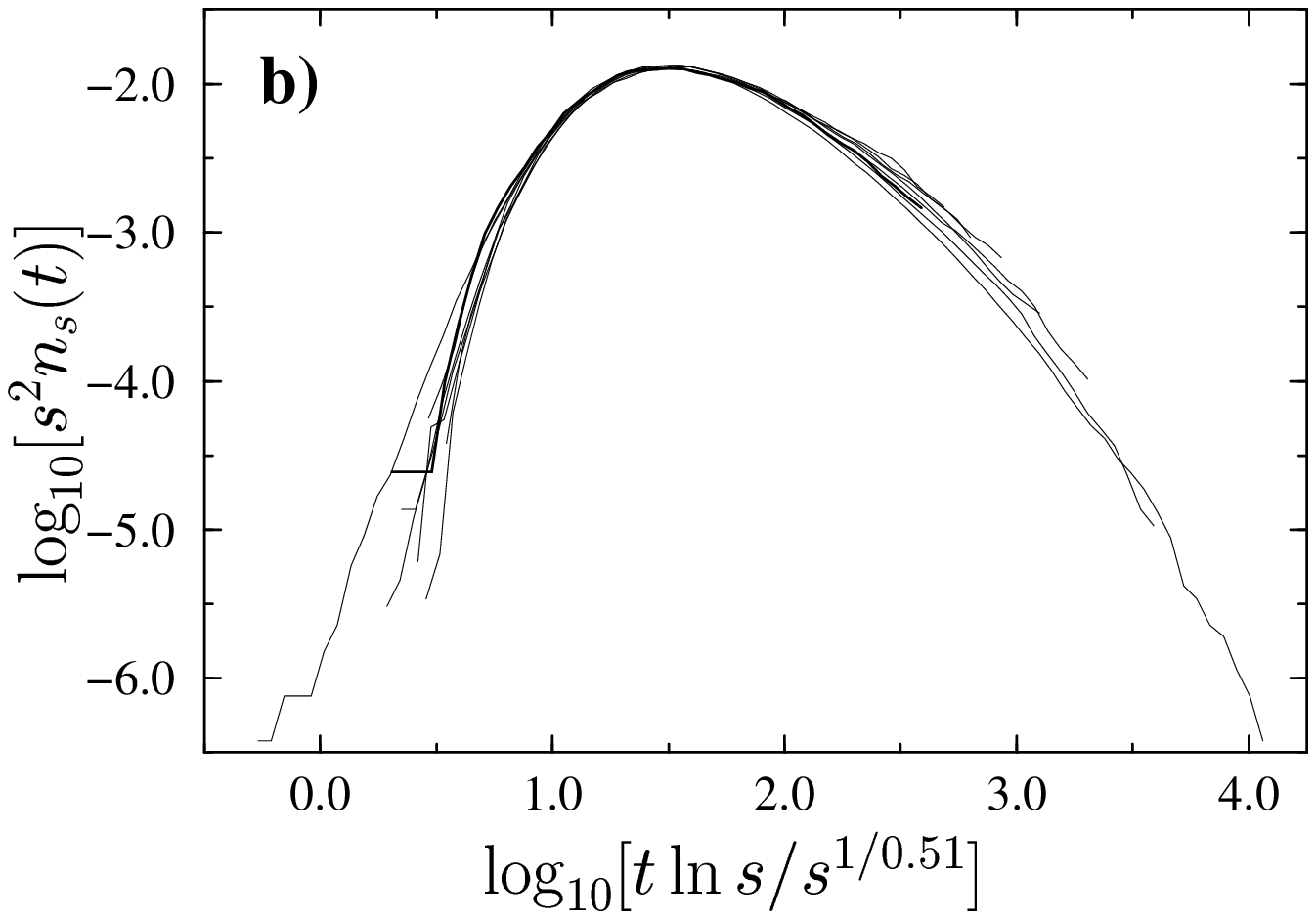,
      bbllx=131, bblly=375, bburx=540, bbury=660, width=8.5cm}}
  \caption{Same data as in the previous figure. (a) Mean
      cluster size as a function of $t\ln S(t)$. (b) Finite-size
      scaling of the cluster size distribution.}
  \label{fig:2d-logcorrect}
\end{figure}

The evidence exposed in the previous figures seems to lend support to
the hypothesis of a simple power law behavior.  However, the situation
is not completely clear, as we show in Figure~\ref{fig:2d-logcorrect}.
Fig.~\ref{fig:2d-logcorrect}(a) tests the theoretical prediction
\equ{eq:mfld}.  Plotting $S(t)$ as a function of $t\ln S$ in
logarithmic scale we obtain a slope $\zeta=0.51$, very close to the
expected value $1/2$ in $d=2$. Moreover, we remark that the
goodness-of-fit in this case, as measured by the Pearson's $r$
coefficient \cite{press97}, is higher than the one obtained from the
linear regression in Fig.~\ref{fig:2d-hidroaniso}(a).  Indeed,
least-squares fittings covering the last three orders of magnitude in
the abscissae of Figs.~\ref{fig:2d-hidroaniso}(a) and
\ref{fig:2d-logcorrect}(a), yield the values $r=0.99960$ and
$r=0.999948$, respectively.  In Fig.~\ref{fig:2d-logcorrect}(b) we
check the finite size scaling hypothesis by plotting $s^{2}n_s(t)$ as
a function of the rescaled time $t\ln s/s^{1/{\zeta}}$. The best
collapse again results from a value of $\zeta=0.51$.

\begin{figure}[t]
  \centerline{\epsfig{file=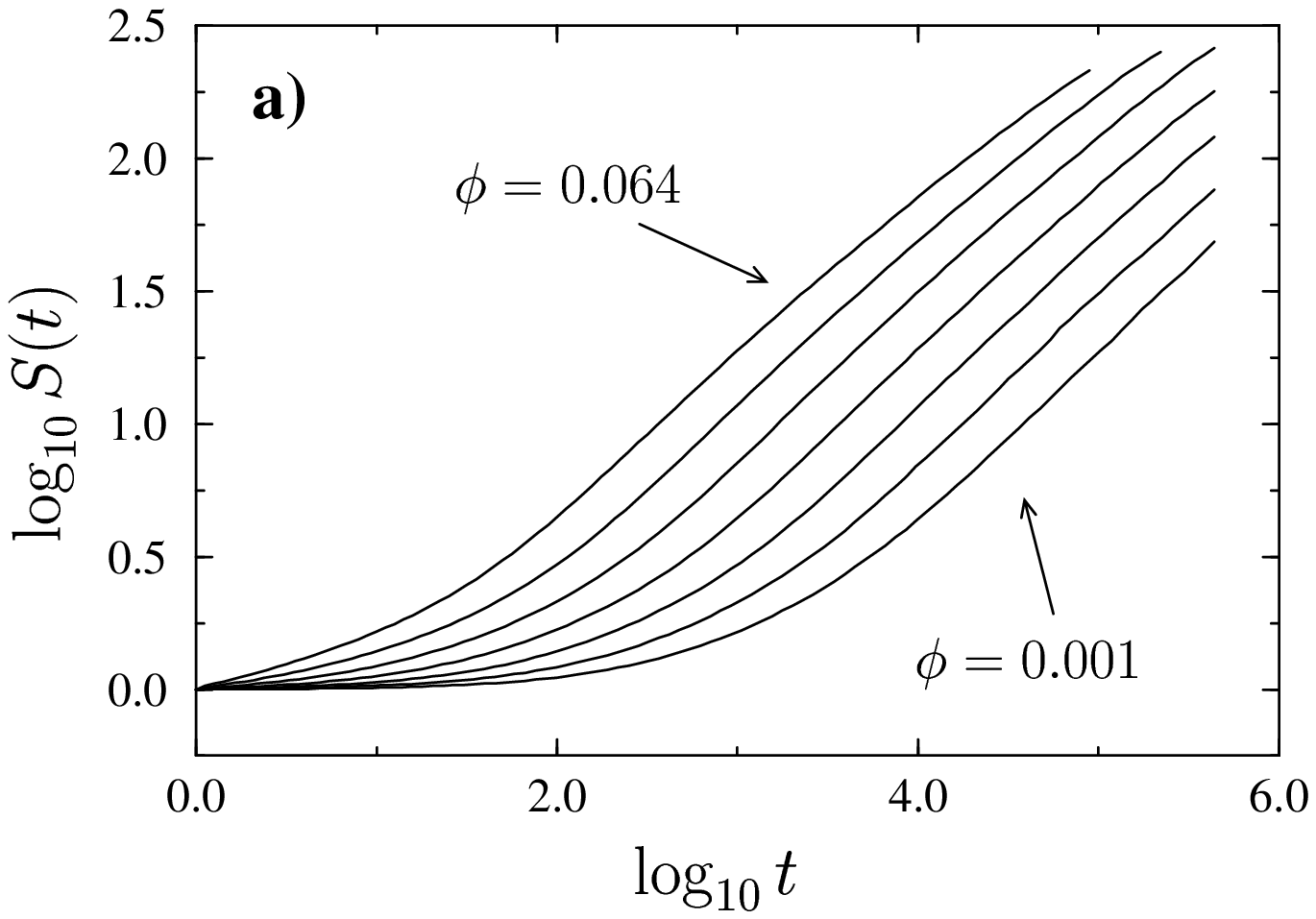,
      bbllx=131, bblly=375, bburx=540, bbury=660, width=8.5cm}}
  \centerline{\epsfig{file=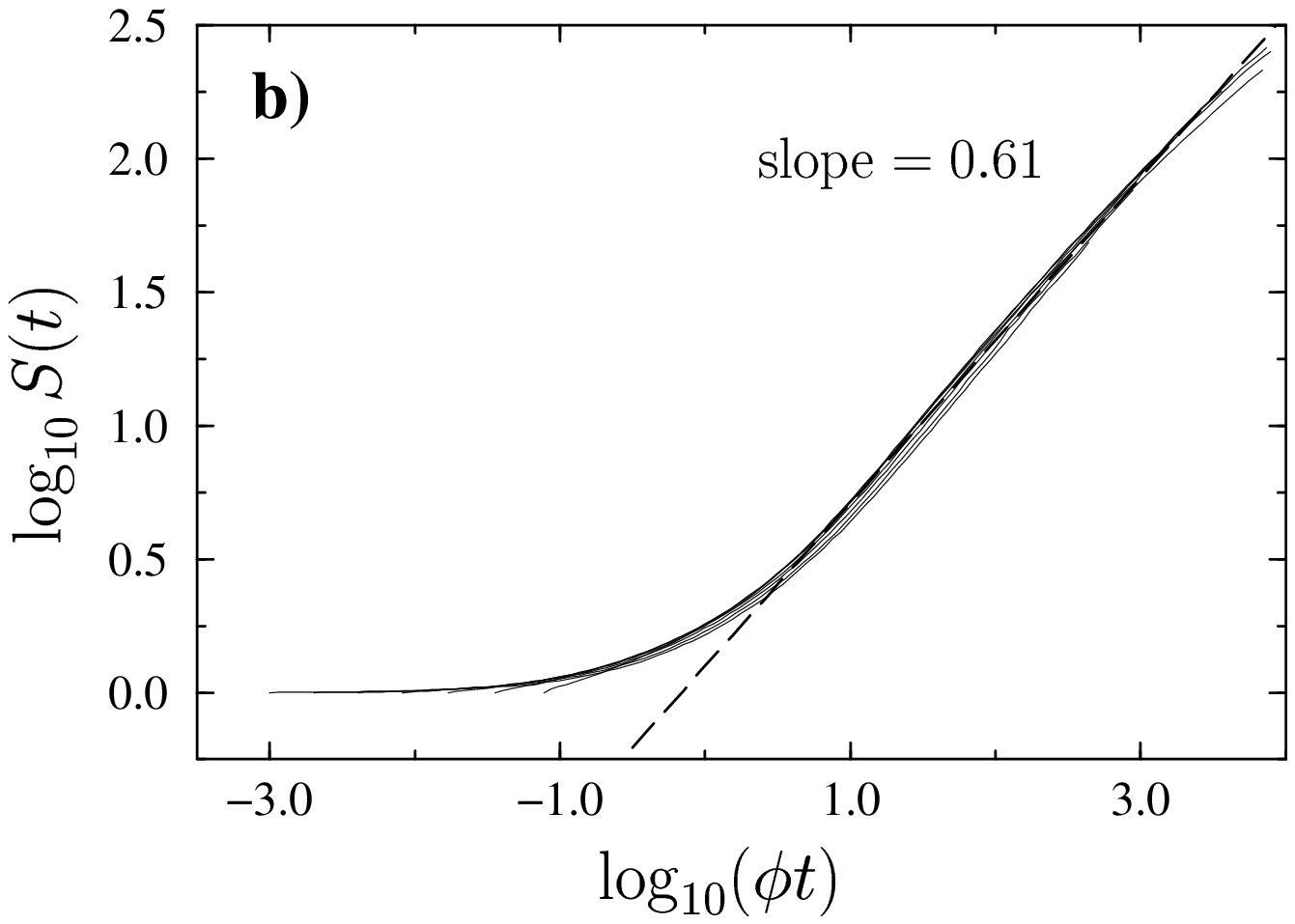,
      bbllx=131, bblly=375, bburx=540, bbury=660, width=8.5cm}}
  \caption{Mean cluster size for anisotropic aggregation, for several
    values of the volume fraction.  From top to bottom, $\phi=0.064,
    0.032, 0.016, 0.008, 0.004, 0.002,$ and $0.001$. (a) As a function
    of time. (b) As a function of the rescaled time $\phi t$. The
    average slope obtained from the common scaling region is $z=0.61$.}
  \label{fig:multiphi}
\end{figure}

\begin{figure}[t]
  \centerline{\epsfig{file=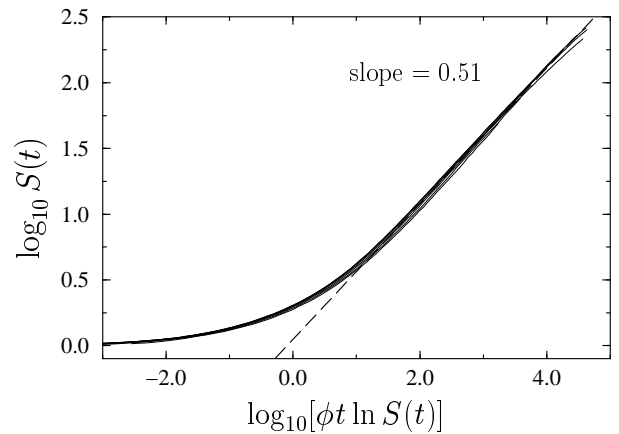,
      bbllx=131, bblly=375, bburx=540, bbury=660, width=8.5cm}}
  \caption{Same data as in the previous figure, now as a function of
      the rescaled quantity $\phi t \ln S$. The
    average slope obtained from the common scaling region is $\zeta=0.51$. }
  \label{fig:multiphilog}
\end{figure}

In the simulations presented above we have been able to match the
observed value of $z$ by selecting the particular value of $\phi$
reported in the experiments. It has been argued, however, that the
value of $z$ may depend on the volume fraction $\phi$
\cite{promislow95}. In order to verify the accuracy of this statement,
we have performed simulations for several values of $\phi$, ranging
from very dilute ($\phi\simeq0.001$) to moderate concentrations
($\phi\simeq0.1$), in a system of fixed size $1024\times512$.
Figure~\ref{fig:multiphi}(a) plots the mean cluster size $S(t)$ as a
function of $t$ for the different volume fractions considered. The
slope of the different graphs depicted in this figure does not seem to
depend on $\phi$.  This fact becomes even more clear in
Fig.~\ref{fig:multiphi}(b), where we have plotted $S(t)$ as a function
of the {\em rescaled time} $\phi t$. We observe that in this case all
the plots {\em collapse} onto a universal function, independent of the
volume fraction. The collapse of the different graphs is also shown in
Fig.~\ref{fig:multiphilog}, now as a function of the rescaled quantity
$\phi t \ln S$. Note that Eq. \equ{eq:mfld} leads directly to the
required scaling factor of $\phi$ in $d=2$. As in Figure
\ref{fig:2d-logcorrect}, the collapse is statistically better in terms
of this new rescaling than for the single power law.

The collapsed plots allow us to select a common scaling region for all
of them, from which we can extract slopes for the different values of
$\phi$ that are directly comparable.  An average slope can thus be
defined.  In the case of Fig.~\ref{fig:multiphi}(b) (single power law
interpretation), individual slopes range between $0.59$ and $0.63$.
From them, we obtain an average exponent $z=0.61$, in accordance with
our previous estimate and the experimental results in
Ref.~\cite{fraden89}. In addition, from Fig.~\ref{fig:multiphilog}
(power law with logarithmic corrections), we obtain slopes between
$0.49$ and $0.53$, yielding an average value of $\zeta=0.51$.

\subsection{Anisotropic diffusion: saturated regime}

\begin{figure}
  \centerline{\epsfig{file=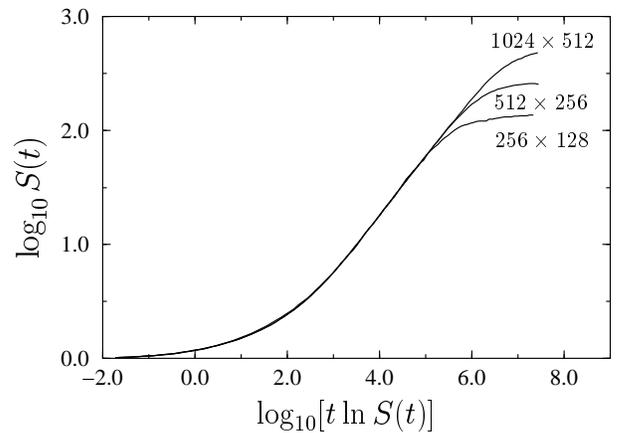,
      bbllx=131, bblly=375, bburx=540, bbury=660, width=8.5cm}}
  \caption{Mean cluster size as a function of $t\ln S(t)$, for
    anisotropic aggregation in the saturated regime. The plots
    correspond to a fixed volume fraction $\phi=0.019$ and different
    system sizes. The onset of the plateau and its height are 
    functions of the system volume.}
  \label{fig:satura1}
\end{figure}

\begin{figure}
  \centerline{\epsfig{file=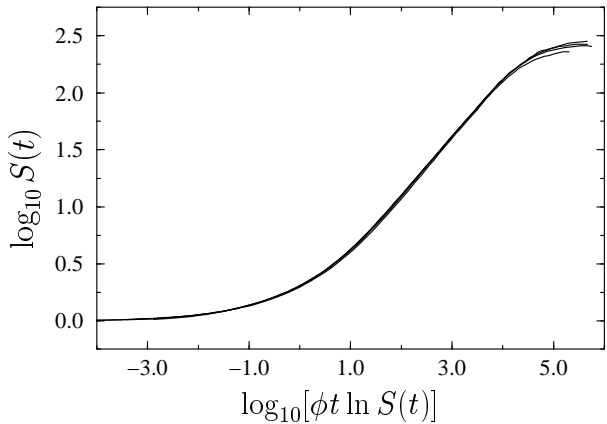,
      bbllx=131, bblly=375, bburx=540, bbury=660, width=8.5cm}}
  \caption{Mean cluster size as a function of the rescaled quantity $\phi
    t\ln S(t)$, for anisotropic aggregation in the saturated regime.
    The plots correspond to a fixed system size $512\times256$ and
    different volume fractions $\phi=0.0048, 0.0095, 0.0191,$ and
    $0.038$. The collapse of the curves shows that both the onset and
    the height of the saturated plateau depend only on the system
    size.}
  \label{fig:satura2}
\end{figure}

In view of the results presented in the last section, we naturally
conclude that the slope does not depend on the volume fraction and
that the disagreement with prior experimental findings
\cite{promislow95} may be a consequence of both logarithmic
corrections and finite-size effects.  In experiments or computer
simulations, finite-size effects can cause the system to crossover, at
large times, to a {\em saturated} regime in which any predicted
scaling behavior is lost.  We have investigated this issue by
simulating systems which were allowed to evolve for very long
times---up to ten times longer than in our previous simulations.
Figure~\ref{fig:satura1} depicts the mean cluster size computed for
different system sizes, while keeping the same fixed initial volume
fraction, $\phi=0.019$. In Fig.~\ref{fig:satura2}, we plot $S(t)$ for
a fixed value of the system size $512\times 256$, and different values
of $\phi$. The chief feature of these plots is the onset of a plateau
whose location and height appears to be a function of the system
volume $V$.  This flatter region is indicative of a considerable
slowing down in the dynamics.  Finite size effects, unavoidable at
relatively large times, can corrupt the interpretation of any expected
scaling behavior.

\subsection{Anisotropic diffusion: quasi one-dimensional  regime}

We have established the existence of deviations from the dilute
asymptotic behavior due to saturation when the system evolves up to
very long times. In addition, in a highly concentrated system, one can
also observe a crossover to a {\em quasi one-dimensional regime}, as
first remarked by Miyazima {\em et al.}  \cite{miyazima87}. At the
late stages of evolution in a very concentrated system, the largest
chains attain a length which is of the order of the system size. Their
effect is therefore to divide the plane into one-dimensional strips.
The aggregation of the remaining smaller clusters is thus mainly
restricted to occur within these strips \cite{fraden89}, becoming
effectively a one-dimensional process.

\begin{figure}[t]
  \centerline{\epsfig{file=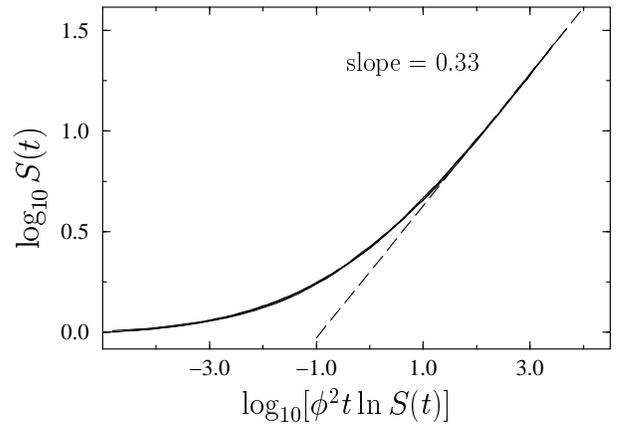,
      bbllx=131, bblly=375, bburx=540, bbury=660, width=8.5cm}}
  \caption{Mean cluster size as a function of the rescaled quantity
    $\phi^2 t\ln S(t)$, for anisotropic aggregation in $d=1$. The
    plots correspond to a fixed system size $10000$ and different
    volume fractions $\phi=0.02, 0.03$, and $0.04$. The average slope
    obtained from the common scaling region is $\zeta=0.33$.}
  \label{fig:1dmodel}
\end{figure}

In a truly one-dimensional model of aggregation, according to Eq.
\equ{eq:mfld}, we expect the mean cluster size to satisfy $S(t)\sim
[\phi^2 t \ln S(t)]^{1/3}$. For the sake of completeness, we have
adapted our model to simulate aggregation within a line. The only
relevant change in the algorithm concerns the way in which the
direction of the tentative movements is selected---either to the right
or to the left, with equal probability. The results of the simulations
are plotted in Fig.~\ref{fig:1dmodel}. This figure corroborates both
the correct scaling of time with $\phi^2$ (in contrast to $\phi$ for
$d\geq 2$), and an exponent $\zeta=0.33$.

\begin{figure}[t]
  \centerline{\epsfig{file=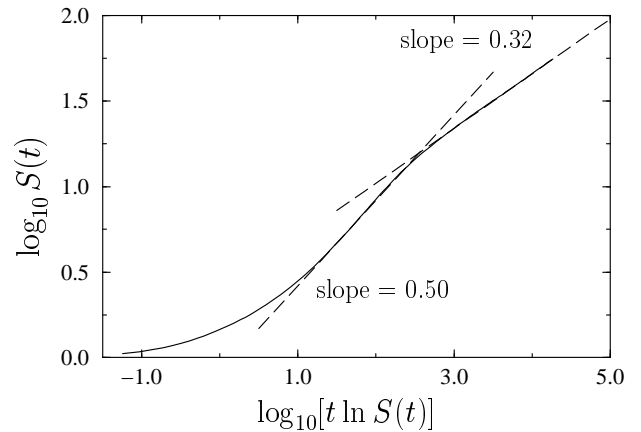,
      bbllx=131, bblly=375, bburx=540, bbury=660, width=8.5cm}}
  \caption{Mean cluster size as a function of $t\ln S(t)$,  for
    anisotropic aggregation in $d=2$, in the quasi one-dimensional
    regime. The dashed lines have slopes $\zeta=0.32$, at larger
    times, and $0.50$ at intermediate times, corresponding to the
    asymptotic dilute regime.}
  \label{fig:crossover}
\end{figure}

We next present results from simulations in $d=2$ of a system
exhibiting quasi one-dimensional behavior. The data in Figure
\ref{fig:crossover} correspond to a volume fraction $\phi=0.48$ in a
system of volume $1024\times512$ \cite{footn2}.  After a transient
regime, we observe that the slope stabilizes in a value $\zeta \simeq
0.32$, clearly smaller than the one found for less concentrated
systems, and in excellent agreement with the exponent $\zeta=1/3$
expected in $d=1$. We note that both the slope and the crossover time
do not depend on the volume of the system for a fixed value of $\phi$.
The results obtained from simulations in a volume $512\times 256$ (not
shown) confirm this statement.

We have further investigated the difference between the saturated and
the quasi one-dimensional regimes. An effective way of doing so is to
consider the movement of the individual clusters. As discussed above,
in the quasi one-dimensional regime, average-sized clusters are
confined to move within the narrow strips delimited by the largest
chains. Therefore, it is more likely for a cluster to move along the
direction parallel to the $Z$-axis, than in any other direction
perpendicular to it. This effect can be quantitatively assessed by
estimating the relative frequency with which clusters move in a given
direction $\theta$. To this end, we define the {\em jump orientation
  density} $F(\theta)$ as follows: At the last stages of the
simulations (usually the last $25\%$ of the time alloted for the run)
we keep track of the direction $\theta$ of all the {\em accepted}
movements performed by the clusters. $F(\theta)d \theta$ is then
defined as the probability that any of those movements is directed
along a direction included in the interval $[\theta, \theta + d
\theta]$. At late times in the evolution of the system, most of the
clusters have a length larger that the cut-off $s_0$. If there is no
bias in the direction taken by the accepted steps, we expect that the
function $F(\theta)$ will match the {\em a priori} angular
distribution \equ{probangle-2d}. In the quasi one-dimensional regime,
on the other hand, given that the clusters move with higher
probability along the $Z$-axis, we expect $F(\theta)$ to show
anomalous maxima at $\theta=0$ and $\theta=\pi$.

\begin{figure}[t]
  \centerline{\epsfig{file=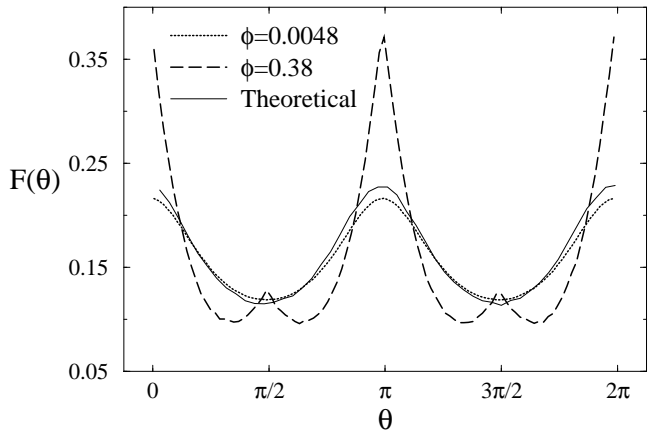, width=8.5cm}}
  \caption{Jump orientation density $F(\theta)$ for two systems of
    size $512\times256$, at two different volume fractions. The
    dilute system $(\phi=0.0048)$ matches the {\em a priori} angle
    distribution, Eq.~\equ{probangle-2d} (full line). The concentrated
    system $(\phi=0.38)$, in the quasi one-dimensional regime, shows
    two sharp peaks at $\theta=0$ and $\theta=\pi$.}
  \label{fig:angles}
\end{figure}

In Figure~\ref{fig:angles} we plot $F(\theta)$ as computed from two
systems of size $512\times256$, with very different values of $\phi$.
The dilute system $(\phi=0.0048)$ is well inside the saturation
regimen. As we can check from this plot, its angular distribution of
jumps matches quite closely the {\em a priory} distribution of
directions, plotted as a reference in full line. In the case of the
highly concentrated system $(\phi=0.38)$, we can observe the two peaks
in $F(\theta)$, characteristic of the quasi one-dimensional regime.
The secondary peaks at $\theta=\pi/2$ and $\theta=3\pi/2$ are a
spurious result of the presence, even at the large time considered, of
some clusters of size $s<s_0$ with an isotropic diffusivity.  If they
are removed from the statistics, the secondary peaks disappear.

\section{Aggregation in \lowercase{$d=3$}}
\label{sec:d=3}

Our two-dimensional model can be easily extended to $d=3$.  In this
case, the random directions of the jumps are sampled according to the
distribution \equ{probangle-3d}. Simulations have been performed for
systems of size $512\times64\times64$. Results correspond to averages
over 100 realizations.

\begin{figure}[t]
  \centerline{\epsfig{file=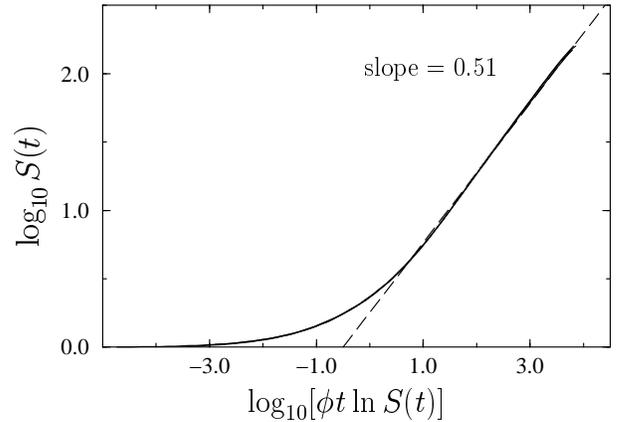,
      bbllx=131, bblly=375, bburx=540, bbury=660, width=8.5cm}}
  \caption{Mean cluster size as a function of the rescaled quantity
    $\phi t \ln S(t)$, for anisotropic aggregation in $d=3$. The volume
    fractions plotted are $\phi=0.0048, 0.024$ and $0.048$.}
  \label{fig:3d-collapse}
\end{figure}

In Fig.~\ref{fig:3d-collapse} we plot our results for the mean cluster
size $S(t)$, for systems with an initial volume fraction $\phi=0.0048,
0.024, 0.048$. The average slope in the scaling region yields an
exponent equal to the one obtained for two-dimensional aggregation,
namely $\zeta=0.51$.  This fact is, however, not surprising: Since we
are above the critical dimension of the problem, $d_c=2$, we expect to
find exponents independent of the dimensionality. As shown in the
figure, the collapse of the mean-cluster size for different volume
fractions as a function of the rescaled time $\phi t \ln S(t)$,
also holds in $d=3$.

\section{Conclusions}
\label{sec:conclu}

In this paper we have investigated the dynamics of the irreversible
aggregation of linear rigid chains oriented along a preferred
direction. A possible physical realization of such system would be the
aggregation of dipolar particles in presence of a strong external
field, oriented along the $Z$ axis. We have proposed a Monte-Carlo
model, whose key ingredient is a corrected anisotropic diffusivity of
the clusters, as expressed in Eq.~\equ{eq:mobilidad}. This diffusivity
exhibits a characteristic logarithmic correction due to hydrodynamic
interactions. A simple heuristic argument suggests that these
logarithmic corrections should also emerge at a macroscopic scale,
reflected in the asymptotic behavior of various quantities such as the
mean cluster size.

We have performed extensive simulations in $d=2$, in a variety of
system sizes and initial volume fractions.  Assuming a simple power
law behavior for the mean cluster size, $S(t) \sim t^z$, our results
yield a dynamic exponent in the scaling region $z\simeq0.61$. This
value is in excellent agreement with previous experimental works
\cite{fraden89}. For dilute systems, the value of the dynamic exponent
seems to be independent of the initial volume fraction.

Alternatively, a better fit of our data is obtained in terms of the
functional relationship $S(t) / [\ln (S(t))]^\zeta \sim t^\zeta$,
which explicitly incorporates logarithmic corrections. Within this
approach, the anomalous value of the dynamic exponent characterizing
the aggregation of anisotropic rod-like clusters is, in fact, the
consequence of the logarithmic corrections superposed to the usual
power law behavior of {\em isotropic} cluster-cluster aggregation.

Long-time evolutions for any value of the volume fraction $\phi$ lead
to a saturated regime, characterized by a drastic slowing-down of the
dynamics. The mean cluster size shows a flat region, whose height and
onset are exclusively functions of the system volume.

In highly concentrated systems, we observe a crossover to a quasi
one-dimensional regime. Characteristic of this transition is a sharp
change in the slope of the mean cluster size, yielding a value
$\zeta=0.32$. To support these findings, we have carried out
simulations in $d=1$, where the mean cluster size exhibits a similar
behavior with $\zeta=1/3$. In the quasi one-dimensional regime, 
clusters have a higher tendency to move along the direction parallel
to their axis, than in any direction perpendicular to it. This point
is made clear by analyzing the jump orientation density $F(\theta)$.

The extension of our model to $d=3$ shows no significant variations
with respect to the case $d=2$. This result is to be expected, given
that the critical dimension of the problem appears to be $d_c=2$.

To conclude, we would like to remark the potential applications of our
model, especially the possibility of characterizing the
aggregation of dipolar colloidal suspensions in different geometries
of technological importance. For example, in a porous medium, due to
the effective reduced dimensionality, we would expect a slowing down of
the aggregation rate, resemblant to the aforementioned quasi
one-dimensional regime.

\acknowledgments

M. C. M. was supported by a grant from the Direcci\'o General de
Recerca (Generalitat de Catalunya) and by the NSF Grant No.
DMR-93-03667. The work of R. P. S. was supported by the Ministerio de
Educaci\'{o}n y Cultura (Spain). We are grateful to Prof.  M. Kardar
for pointing out the possibility of logarithmic corrections in the
scaling of the mean cluster size. We appreciate his critical reading
of the manuscript.



\begin{references}
\bibitem{footn3} Note that, at constant temperature, the diffusion
  coefficient and the mobility of a cluster are proportional. We will
  therefore use both terms indistinctly to refer to $D(s)$.
\bibitem{footn} Note that in the case of a chain, length and mass are
  equivalent magnitudes.
\bibitem{footn2} In these conditions, the assumption of chains being
  the only cluster forming in the system could be unrealistic. At high
  concentrations, far more complicated structures (for instance,
  columns) may also be present.
\end{references}
\end{document}